\documentstyle[prl,aps,epsf]{revtex}
\draft
\epsfclipon

\begin{document}

\twocolumn[\hsize\textwidth\columnwidth\hsize\csname 
@twocolumnfalse\endcsname
\title{Subgap anomaly and above-energy-gap structure 
in chains of diffusive SNS junctions}

\author{T.~I.~Baturina$^+$\/\thanks{e-mail: tatbat@isp.nsc.ru},
D.~R.~Islamov$^{+*}$, Z.~D.~Kvon$^+$}

\address{$^+$Institute of Semiconductor Physics, 630090 Novosibirsk,
Russia\\
$^*$Novosibirsk State University, 630090 Novosibirsk, Russia}

\maketitle

\begin{abstract}
We present the results of 
low-temperature transport measurements on chains 
of superconductor--normal-constriction--superconductor (SNS) junctions
fabricated on the basis of superconducting PtSi film. 
A comparative study of the properties of the chains, 
consisting of 3 and 20 SNS junctions in series, 
and single SNS junctions reveals essential distinctions 
in the behavior of the current-voltage characteristics of the systems: 
(i) the gradual decrease of the effective suppression voltage 
for the excess conductivity observed at zero bias as the quantity 
of the SNS junctions increases, 
(ii) a rich fine structure on the dependences 
$dV/dI-V$ at dc bias voltages higher than the superconducting gap and 
corresponding to some multiples of $2\Delta/e$. 
A model to explain this above-energy-gap 
structure based on energy relaxation of electron 
via Cooper-pair-breaking in superconducting island 
connecting normal metal electrods is proposed.

\end{abstract}


]
\bigskip
\narrowtext

In the past few years, the mesoscopic systems, consisting of 
a normal metal (N) or heavily doped semiconductor being 
in contact with a superconductor (S), have attracted an increased
interest mainly because of the richness of the involved quantum 
effects \cite{Pannetier}. 
The key mechanism governing the carrier transport through the NS contact 
is the Andreev reflection \cite{Andreev}.  
In this process, an electron-like excitation 
with an energy $\epsilon$ smaller than
the superconducting gap $\Delta$ moving from
the normal metal to the NS interface is retro-reflected as a 
hole-like excitation, while a Cooper pair is 
transmitted into the superconductor. 
This phenomenon is the basis of the proximity effect, 
which generally implies the influence of a 
superconductor on the properties of 
a normal metal when being in electrical contact.
The consequences of Andreev reflection on the current-voltage 
characteristics (CVC) of a NS junction are studied in detail in the
so-called BTK theory \cite{BTK}, which describes the subgap current
transport in terms of ballistic propagation of quasiclassical
electrons through the normal-metal region, accompanied by Andreev
and normal reflections from NS interface.
The probability of Andreev reflection and normal reflection
are energy-dependent quantities and
the relation between them depends on the barrier strength at
NS interface, that is characterized by a parameter $Z$ ranging from $0$ 
for a perfect metallic contact to $\infty$ for a low transparency tunnel
barrier. For a perfect contact ($Z=0$), probability of
Andreev reflection is equal to unity for particles with an energy 
$\epsilon$ smaller than the superconducting gap $\Delta$ and
the subgap conductance are found to be twice the normal state 
conductance thus demonstrating the double charge transfer. 
In the other limit, when $Z \to \infty$, the conductivity is very small.

When a normal metal is placed between two superconducting electrods
another mechanism is involved in charge transfer. It is multiple
Andreev reflection process (MAR). The concept of MAR was first 
introduced by Klapwijk, Blonder and Tinkham \cite{KBT}
in order to explain the subharmonic energy gap structure
observed as dips in the differential resistance
$dV/dI$ of SNS junctions at the voltages 
$V_n=2\Delta/ne$ ($n=1,2,3,\cdots$).
In all systems, diffusive as well as
ballistic, the MAR mechanism relies on the quasiparticles
to be able to add up energy gained from multiple passages.
This energy gain results in strong quasiparticle nonequilibrium.
Calculations of the CVC, within the approach developed in Ref.
\cite{BTK}, and taking into account MAR process was performed
in \cite{OTBK} (OTBK theory).
It was shown that MAR process results not only in SGS at voltages 
$V_n=2\Delta/ne$ but affects the general form of CVC at all voltages 
as well. 
Although BTK theory is not intended to describe diffusion
transport in normal-metal region, a suitable value $Z\sim 0.55$ of the
barrier strenght gives result very close to those obtained
by Green's-function method for microconstriction in the dirty limit
\cite{AVZ}. Experimentally, in some cases 
BTK theory has been successful in providing
a method to extract the junction reflection coefficients based on
experimental CVC \cite{Huf,Kutch1}. 

At present, a large variety of SNS junctions
(most of them are diffusive) is being fabricated and studied
\cite{Huf,Kutch1,Cour,Char,Dynes,Hoss,PRB00,PRB01}.
The investigations of these junctions are primarily focused
on the nonlinear behavior of the current-voltage characteristics,
which exhibit the anomalous resistance dip at zero bias 
[zero bias anomaly (ZBA)],
the subharmonic energy gap structure (SGS), etc. 
These experiments have revealed a number of peculiar properties
of diffusive SNS junctions, the explanation of those being beyond
the ballistic BTK and OTBK theories. 
Nowadays the properties of diffusive SNS junctions
are considered to be determined not only by  
the parameters of the NS interface, but   
nonlocal coherent effects as well,
namely (i)~the superposition of multiple coherent scattering
at the NS interfaces in the presence of disorder
(so-called reflectionless tunneling \cite{RT}) and 
(ii)~electron-electron interaction in the
normal part.
The latter is one of the important points of a recently 
developed ``circuit theory"
when applied to diffusive superconducting hybrid systems\cite{NazarovStoof}.
Within this approach, which is based 
on the use of nonequilibrium Green's functions,
the electron-electron interaction induces a
weak pair potential in the normal metal. It results not only in a change of 
the resistance, but also in a non-trivial distribution of the electrostatic
and chemical potentials in the structure, which implies non-local
resistivity.  
Following the spirit of Nazarov's circuit theory Bezuglyi {\it et al}.
\cite{Bez} have developed a consistent theory of carrier transport
in long diffusive SNS junctions with arbitrary transparency of NS
interfaces. 
Although much work both theoretical and experimental
has been done on single SNS junctions, 
it is a challenge to fabricate and carry out 
comparative measurements on multiply connected SNS systems. 
In this paper, we present the results of low-temperature transport 
measurements on chains of SNS junctions 
and on single SNS junctions and perform the comparative analysis 
of their properties.

The design of our samples is based on the fabrication technique
of SNS junctions that has been recently proposed and realized 
by us for preparation of single SNS junctions and two-dimensional
arrays of SNS junctions \cite{PRB00,PRB01}.
The main idea employed in those experiments was to use
superconductive and normal region made of the same material.
The point is the suppression of superconductivity
in the submicron constrictions made in an ultrathin 
polycrystalline PtSi superconducting film 
by means of electron beam lithography and subsequent plasma etching. 
Here we use the same fabrication technique to prepare chains of
SNS junctions. 

The original PtSi film (thickness -- 6~nm) was formed on a Si substrate.
The film was characterized using Hall bridges
$50~\mu$m wide and $100~\mu$m long.
The film had a critical temperature $T_{\mathrm{c}}=0.56$~K.
The resistance per square was 104~$\Omega$.
The carrier density obtained
from Hall measurements was $7\cdot10^{22}$~cm$^{-3}$,
corresponding to a mean-free path $l=1.2$~nm and a diffusion constant
$D=6$~cm$^2$/s, estimated using the simple free-electron model.

Single SNS junction is a constriction between two holes made in the film
and placed in the Hall bridge.
The hole diameters are $1.7~\mu$m and 
the distance between centres of holes is $2.1~\mu$m,
resulting in a width of narrowest part of the constriction 
of $0.4~\mu$m.
A scanning electron micrograph and a schematic view
of a sample are presented in Fig.1a,b.
The chains of SNS junctions are a series of constrictions connected
by the islands of the film (Fig.1c,d). They are designed to provide 
the possibility of the comparative study. 
The dimensions of constrictions are identical with 
those of the single SNS junction, and the characteristic
size of the islands of the film is $\sim 1.3~\mu$m.
As the constrictions are not superconducting 
we have a chain of SNS junctions (Fig.1e).

The measurements were carried out with the use of a phase sensitive
detection technique at a frequency of 10~Hz that allowed 
us to measure the differential resistance ($dV/dI$)
as a function of the dc current ($I$).
The ac current was equal to 1--10~nA.
Figure~2 shows typical dependences of $dV/dI$-$V$ for 
the samples with single constriction (1D-1) and chains,
consisting of 3 (sample 1D-3) and 20 (sample 1D-20) 
SNS junctions in series.
On the abscissa the value of voltage obtained by numerical integration 
of the experimental dependences $dV/dI$-$I$ and afterward divided by
the number of SNS junctions is plotted.
Measured differential resistance is also divided by
the number of SNS junctions (by 3 for structure 1D-3 and by 20
for structure 1D-20).
Thus for the chains data presented in Fig.2 are dependences
the differential resistance as a function of dc bias voltage 
per SNS junction in the average.
As is seen at high voltage for chains the average resistances 
of SNS unit of chains obtained in such a way 
are close to each other and to the value $dV/dI$
of the structure with single SNS junction.
A common feature for all structures studied is 
a minimum of the differential resistance at zero bias voltage.
It points out the high transparency of NS interfaces in these samples.
The data for single SNS junction (1D-1) exhibit a behavior 
very similar to that reported in our previous work \cite{PRB00}.
In Ref.~\cite{Bez} the differential resistance of long diffusive SNS
junction with perfect NS interfaces at zero voltage was estimated by
the following expression:
\begin{equation}
dV/dI(0)=R_N(1-2.64 \xi _N/L),
\end{equation}
where $\xi _N=\sqrt{\hbar D /2 \pi kT}$ is the decay lenght for the 
pair amplitude, $L$ is the lenght of the normal-metal region, and
$R_N$ is its resistance at $T>T_{\mathrm{c}}$.
The expression (1) has obtained on condition that
the inelastic lenght $l_\epsilon$ exceeds the junction lenght $L$.
The energy relaxation is described 
by the time $\tau _\epsilon ^{-1} = \pi \epsilon ^2/(8 \hbar E_F)$.
This is the time it takes for a ``hot" quasiparticle with energy 
$\epsilon$ much larger than temperature $T$ to thermalize with all
the other electrons.
For the excitation energy $\epsilon = 2 \Delta$ 
($\Delta/e \sim 270$~$\mu$V)
we obtain  $l_\epsilon = \sqrt{D \tau _\epsilon} \sim 4.6 $~$\mu$m.
The magnitude of $l_\epsilon$ really exceeds the characteristic
lenght of the constriction.
As $R_N$ we take the difference between the resistance of the whole 
structure with constriction and the resistance of the original film 
without constriction measured between the same probes
at $T>T_{\mathrm{c}}$.
It gives $R_N \sim 530$~$\Omega$. 
Assuming $L \sim 1$~$\mu$m, we find from (1) at $T=100$~mK 
the differential resistance at zero bias 
$dV/dI(0) \sim 0.8R_N =420$~$\Omega$. 
As is seen in Fig.2, this value is close fit
to measured $dV/dI(0)$ for the sample with single constriction 1D-1. 
A clear-cut distinction between dependences $dV/dI$-$V$
of the single SNS junction and the chains consists in 
non-monotone behavior of $dV/dI$-$V$ characteristics of the latter.
The differential resistance of the chains has a minimum at zero bias
voltage and shows a maximum at a finite bias voltage of about 
250~$\mu$V for the sample 1D-3 and 30~$\mu$V for 1D-20.
The second feature is the gradual decrease of the effective 
suppression voltage for the excess conductivity observed at zero 
bias as the quantity of the SNS junctions increases.
The same behavior -- non-monotonic $dV/dI$-$V$ characteristics and
the decrease of the effective suppression voltage for the excess conductivity
in comparison of that in single SNS junctions --
has been observed in two-dimensional arrays of SNS junctions \cite{PRB01}.

The most unexpected result obtained for chains of SNS junction is
the presence of a fine and fully symmetrical structure
as dips in the differential resistance at high biases (Fig.3a,b). 
The positions of these dips approximately correspond to 4, 6, 
and 12 multiples of the superconducting gap. 
As is seen in Fig.3c,
the temperature dependence of the positions of these dips
actually reflects the dependence $\Delta(T)$. 
To our knowledge the results presented in this paper
is the first investigation of chains of SNS junctions
and the first observation such above energy gap structure
(AGS). To offer an explanation of the AGS let us consider 
a current driven in the normal region at high voltages.
As the inelastic lenght more than the normal metal region, 
at finite voltage quasiparticle passing through N$_1$ 
between two superconductors S$_1$ and S$_2$ (Fig.4b,d) results in a 
nonequilibrium energy distribution of quasiparticles.
For a start we address this problem to the 
ballistic OTBK theory \cite{OTBK}.
In this approach the quasiparticles
are divided into two subpopulation which depend on their direction
of motion $f_\to (E)$ and $f_\gets (E)$, with all energies measured with
respect to the local chemical potential.
The current through the junction is 
\begin{equation}
I \propto \int_{-\infty}^{+\infty}[f_\to (E)-f_\gets (E)]dE.
\end{equation}
In Fig.4a,c we have plotted $f_\to (E)$, $f_\gets (E)$,
and $[f_\to (E)-f_\gets (E)]$, assuming a parameter $Z=0.55$.
The voltage across the junction equals $4 \Delta /e$ 
and $6 \Delta /e$ in Fig.4a and Fig.4c, respectively.
The function $[f_\to (E)-f_\gets (E)]$ is depicted by 
the blackness (high values are black) in Fig.4b,d, with
the energy scale being plotted vertically.
These calculations clearly show that the   
nonequilibrium distributions are sharply peacked at the four
gap edges. What happens to quasiparticle injected to 
the superconducting electrode S$_2$
and having the energy more than $2 \Delta$ measured with
respect to the gap edge of S$_2$?
They can decay under spontaneous phonon emission 
into states of lower energy \cite{Eisen}.
Phonons emitted in the process of energy relaxation of injected
quasiparticles can be reabsorbed via Cooper-pair-breaking.
Depending on the volume of finite states, 
the probability of this process is non-monotone function 
of the primary energy of injected quasiparticles. 
It takes the peak values at the energies multipled to $2\Delta$, 
for density of quasiparticle states to have the singularity at the gap edge.
As a result one quasiparticle injected to S$_2$ with energy high the
gap edge by $m \cdot 2 \Delta$ can cause up to $1+2m$ quasiparticles 
injected to N$_2$ (Fig.4b,d) and consequently the increase of the differential
resistance.
The realization of this scenario requires the lenght
of energy relaxation to be comparable 
to dimensions of superconductor. It is really fulfilled in the
SNS systems studied. 
Proposed mechanism supposes the sistem consisting of two SNS junctions
in series to be sufficient in order to observe the AGS. 

In summary, we performed the first comparative study
of the properties of the chains and single SNS junctions.
Our experiments reveal essential distinctions 
in the behavior of the current-voltage characteristics of these systems.
A detailed quantitative analysis needs to take into account
the contributions of nonlocal coherent transport 
and an effect of quasiparticle injection in both the
superconducting and normal regions of chains of the SNS junctions.

We thank R.~Donaton and M.~R.~Baklanov (IMEC) for
providing us the PtSi films, and A.~E.~Plotnikov for the 
performing of electron lithography.
We acknowledge useful discussions with V.~F.~Gantmakher, V.~V.~Ryazanov, 
Ya.~G.~Ponomarev, and M.~Feigel'man. 
This work has been supported by the program 
``Low-dimensional and mesoscopic condenced systems" 
of the Russian Ministry of Science, Industry and Technology
and by RFBR Grant No. 00-02-17965.

\begin{figure}
\centerline{\epsfxsize3in\epsfbox{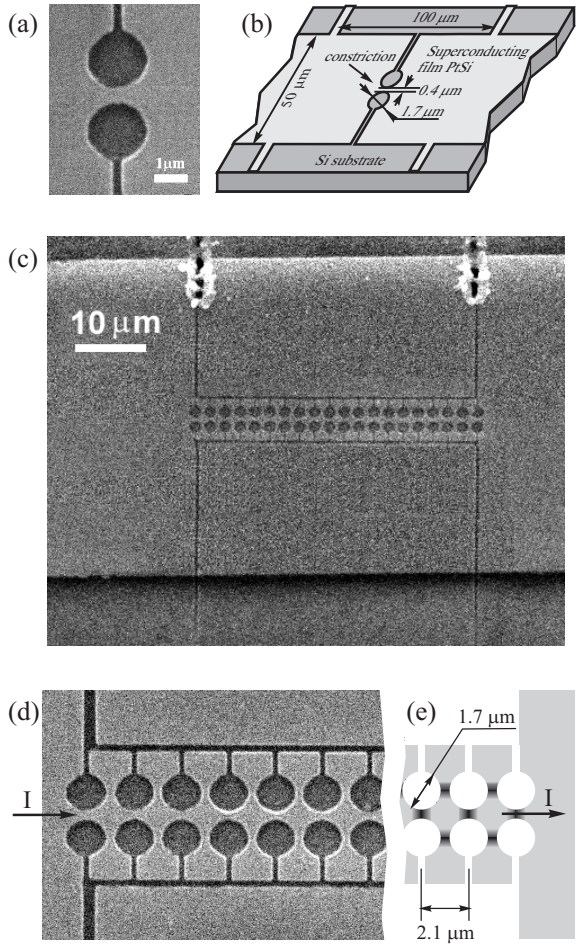}\bigskip}
\caption{(a)~Scanning electron micrograph of the sample with a single 
constriction formed by electron beam
lithography and subsequent plasma etching of the 6~nm PtSi film
grown on a Si substrate.
(b)~Schematic view of a junction (not to scale) showing the layout of the 
constriction in the Hall bridge.
(c)~Scanning electron micrograph of the sample with 20 constrictions.
(d)~SEM subimage of the sample represented in (c).
(e)~The layout of a chain of SNS junctions
showing the dimensions of the structure.
Regions of the normal metal constrictions are dark,
and the superconducting islands are light gray.
}
\label{fig1}
\end{figure}

\begin{figure}
\centerline{\epsfxsize3in\epsfbox{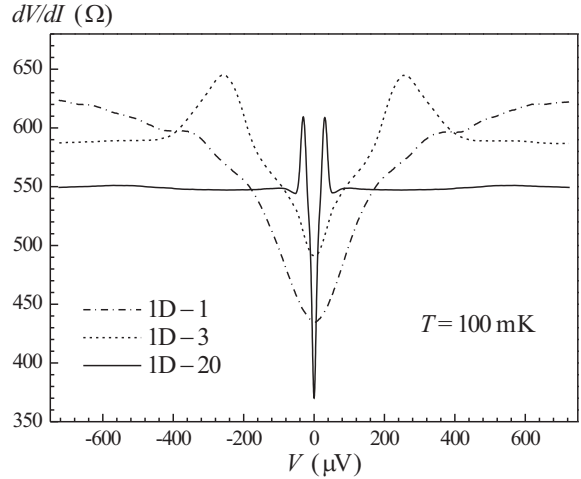}\bigskip}
\caption{Differential resistance vs dc bias voltage for 
samples: (1D-1) with single constriction,
(1D-3) with three constrictions in series,
and (1D-20) with twenty constrictions in series.
$T=100$~mK.
Differential resistance and dc bias voltage
are divided by 3 and 20 for samples 1D-3 and 1D-20,
respectively.
}
\label{fig2}
\end{figure}

\begin{figure}
\centerline{\epsfxsize3in\epsfbox{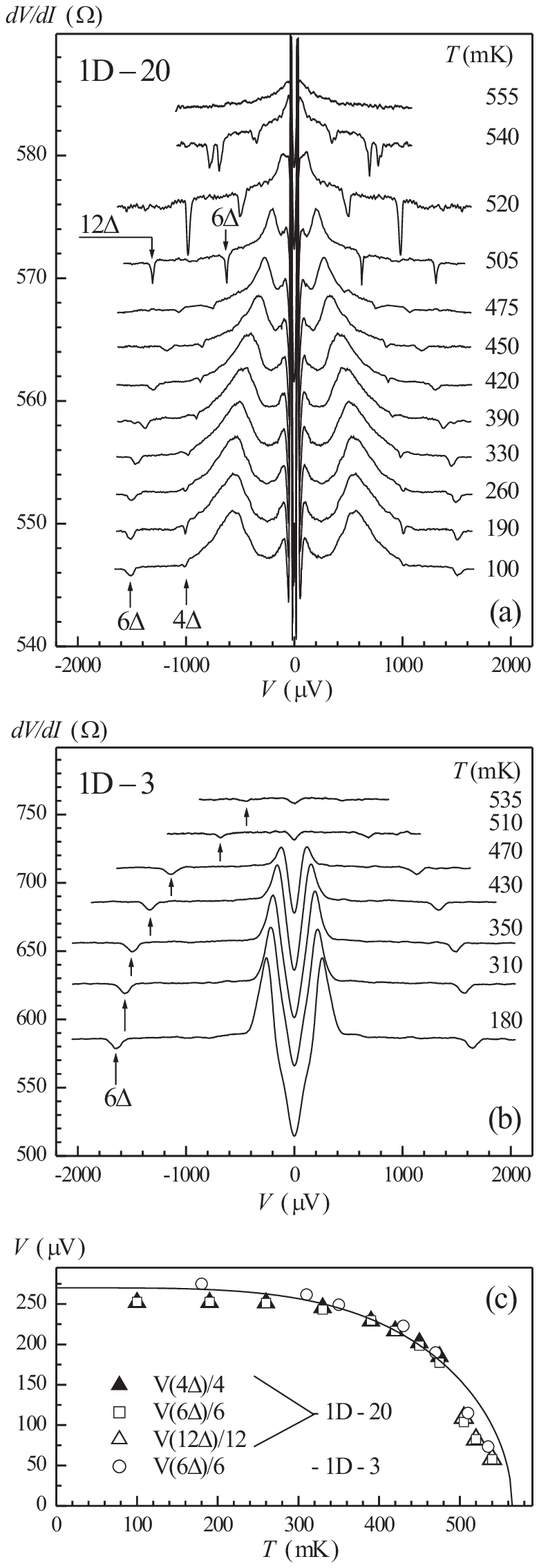}\bigskip}
\caption{(a)~Temperature evolution of the differential resistance 
of the sample 1D-20 as a function of the average bias voltage 
falling at one SNS junction.
All traces except the lowest trace are shifted up for clarity.
(b)~The same for the sample 1D-3.
The differential resistance reveals symmetrical minima
at voltages exceeding $2\Delta$.
The arrows indicate the above-energy-gap structure (AGS),
corresponding to integer multiples of the $2\Delta$.
The temperature dependence of the 
AGS at $eV=m \cdot 2\Delta$ divided by $m$ for both samples 
is depicted in (c) by symbols, and compared to the BCS 
temperature dependence of the gap (solid line).
}
\label{fig3}
\end{figure}

\begin{figure}
\centerline{\epsfxsize3in\epsfbox{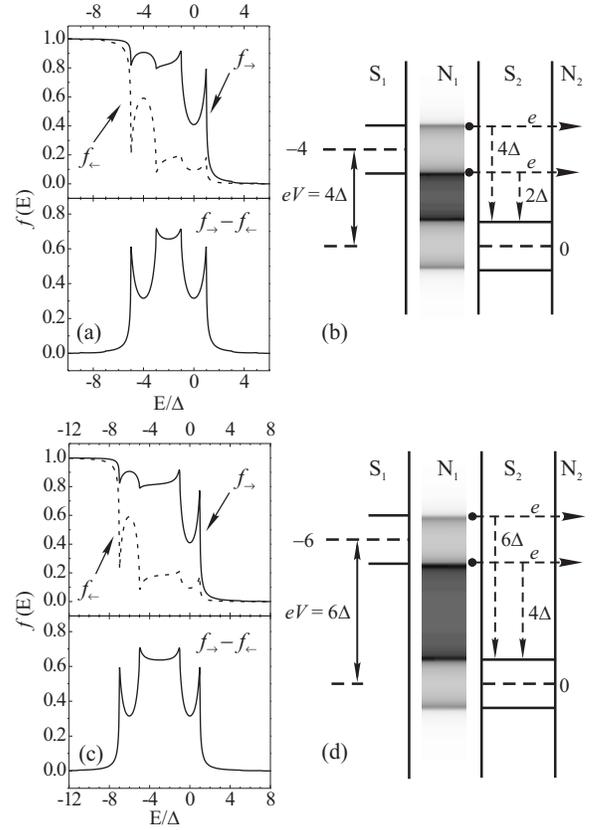}\bigskip}
\caption{Nonequilibrium distributions in normal region, for SNS junction
with $Z=0$: (a,b)~at $eV=4 \Delta$; (c,d)~at $eV=6 \Delta$ and $T=0$.
In (b) and (d) the function $[f_\to (E)-f_\gets (E)]$ is depicted by 
the blackness (high values are black). The energy scale is plotted 
vertically.}
\label{fig4}
\end{figure}

\end{document}